\documentclass[epsfig,12pt,onecolumn]{article}
\usepackage{amsfonts}
\usepackage{amssymb}
\usepackage{amsmath}
\usepackage{multicol}
\usepackage{graphicx}
\usepackage{float}
\usepackage{caption}
\usepackage[top=2cm,left=2cm,right=2.5cm,bottom=5.5cm,includeheadfoot]{geometry}

\input{tcilatex}

\begin{document}

\markboth{M. BOUSDER, Z. SAKHI, M. BENNAI}

\title{\textbf{A new unified model of dark matter and dark energy in 5-dimensional $%
f(R,\phi )$ gravity}}

\author{M. Bousder$^{1}$\thanks{mostafa.bousder@gmail.com}, Z. Sakhi$^{1,2}$\thanks{Zb.sakhi@gmail.com} , M. Bennai$^{1,2}$\thanks{mohamed.bennai@univh2c.ma} \\
%EndAName
$^{1}${\small Lab of High Energy Physics, Modeling and Simulations,}\\
\ {\small Faculty of Science,University Mohammed V-Agdal, Rabat, Morocco}\\
$^{2}${\small Quantum Physics and Applications Team, LPMC, Faculty of
Science Ben M'sik,}\\
{\small Casablanca Hassan II University, Morocco}}

\maketitle

\begin{abstract}
We propose a new unified model that describes~dark energy and dark matter in
the context of $f(R,\phi )$ gravity using a massive scalar field in five
dimensions. The scalar field is considered in the bulk that surrounds the
3-brane in branworld model. We show that for a specific choice of the $%
f(R,\phi )$ function, we can describe the Einstein gravitation in
4-dimensional space-time. We obtain a relationship between the speed of the
universe's expansion and the speed of the bulk's expansion. We also propose
that the dark matter is represented by the scalar field mass and that the
dark energy is a kinetic energy of this field. Finally, we show that,
according to conditions, one can obtain the percentages of density\ of dark
matter and the density of ordinary matter.
\end{abstract}

\section{Introduction}

Recently, various models of modified gravity have been proposed in order to
solve the problem of dark energy (DE) and dark matter (DM) \cite{R1}, using
a function $f(R)$ which depends essentially on the Ricci scalar $R$ \cite{R2}%
. Various models of f(R) modified gravity were proposed, as $R^{2}$
inflation \cite{R3}, which can desribe the cosmic acceleration of the
universe and also inflation, the Brans-Dicke theory, depending on the
coupling constant $\omega $ \cite{R4}, which is a model with a scalar field
describing the gravitation, or the $1/R$ gravity \cite{R5} which can explain
the acceleration of the Universe's expansion.\newline
There are several ways to choose the form of the function $f(R)$. Note that,
only a few models that can describe dark energy, dark matter or inflation,
and conditions for the cosmological viability of dark energy models were
introduced \cite{R6}. Recall that the modified gravity theories are
considered as a perturbation theories around the Ricci scalar in the context
of the Jordane frame or depending on a Ricci scalar and a new scalar field $%
\phi $ coupling in the Einstein frame \cite{R1,4}. \newline
On the other hand, there are other more generalized\ modified gravity
depending on a scalar field whose the coupling with the Ricci scalar is
described by a certain function $f(R,\phi )$ \cite{71,711}. While the other
models of $f(R)$-brane \cite{73} or $f(R,\phi )$-brane \cite{7}, describe
modified gravity in the braneworld scenario in five-dimensional space-time
\cite{10}.

In the present work, we propose a new unified model that describes~dark
energy and dark matter in the context of $f(R,\phi )$ gravity using a
massive scalar field in five dimensions. The scalar field is considered in a
bulk that surrounds the 3-brane in branworld model. We study the evolution
of a scalar fields $\phi $ in modified gravity in 5-dimensional braneworld
model. We have supposed that DM and DE are only the interactions of the
field $\phi $ and the 3-brane \cite{9}. For this reason, we will describe
the modified gravity in 4-dimensional by the braneworld models \cite{13}.
Recall that, the scalar field density on 3-brane depends on the curvature of
space-time.

In the second section, we recall some aspects of the $f(R)$ modified gravity
and introduce a particular form of $f(R,\phi )$ gravity, with a
non-minimally coupling of the Ricci scalar $R$ to the scalar field $\phi $.
We study the potential field $\phi $ by using a rotation transformation. In
the third section, we present the braneworld $f(R)$ modified gravity and we
derive, analytically, the field $\phi $ equations in 5-dimensions to obtain
the evolution equation of the field $\phi $. We focus mainly on the
solutions in 5-dimensional braneworld model, by the integration on the
angular quantity $R\phi $. The solutions of this equation give as the
density of the ordinary matter OM and DM. In the section four, we will
compare the field $\phi $ mass with the scalaron mass \cite{SCA}. We analyze
the density of matter in chronology of the Universe. In section five, a
unified description of the DE and DM and several models of modified gravity
have been indroduced. This unification can be realized by using the
extra-dimensions \cite{10, 2}, or by modification of a 4-dimensional
Einstein gravitational theory \cite{MG}. We show that the field $\phi $
describes both the accelerating expansion of the Universe and the dark
matter. However, for the some reasons, we have shown that the bulk is also
expanding. Thus, the expansion of the Universe depends on bulk expansion.
The DE is then defined as the variation of a large values of the scalar
field $\phi $ and the DM is described as a minimal values of the field $\phi
$. Note that, in the ($\Lambda CDM$) cosmological model, the energy density
in the present Universe of OM and DM, that we obtain, is in good agreement
with the Planck data \cite{6}. The last section is devoted to conclusion.

\section{$f(R,\protect\phi )$ modified gravity}

We propose a description of the field $\phi $ in 5-dimensional (the bulk)
and a description of Ricci scalar $R$ at 4-dimensional (3-brane), such that
the curvature of 3-brane depends on the variation of $R$ and $\phi $. If we
propose that field $\phi $ describe also the gravitation, the large
extra-dimension is purely gravitational. We start with the description of
the gravitation on the 3-brane scale, the potential on\ the 3-brane is then
described by the expression \cite{R1}%
\begin{equation}
V(\phi ,R)=\frac{M_{P}^{2}}{2F(R,\phi )}(R-\frac{f(R,\phi )}{F(R,\phi )}),
\label{030}
\end{equation}%
where the function $F(R,\phi )=\frac{df(R,\text{ }\phi )}{dR}$.\newline
Note that, in the Einstein frame, the energy is conserved and can be
transfered by the field $\phi $ outside the 3-brane, and can not be
physically observed. To solve this problem, we propose to introduce a more
general function $f(R,\phi )$ depending on the two fields $\phi $ and $R.$
This will allow us to explain simultaneously the DM, the DE and the Einstein
gravitation. We chose the function $f(R,\phi )$ of the following form: $%
f(R,\phi )=Re^{-\phi R}$. This function represents the Ricci scalar by an
expanential factor, which will describe the perturbation around the
curvature $R$, describing both DM and DE by the following action%
\begin{equation}
I:=\frac{M_{P}^{2}}{2}\int d^{4}x\sqrt{-g}Re^{-\phi R}+I_{m},  \label{3.1}
\end{equation}%
where $I_{m}$ is the matter action.\newline
To find the expression of $F(R,\phi )$ in the Einstein frame, we replace the
Ricci scalar by a constant value $R_{p}=\frac{2}{\sqrt{6}}\frac{1}{M_{p}},$
which is equivalent to $F(\phi )=e^{-\phi R_{p}}$, which is equivalent with
the choice of the conformal transformation for $f(R)$ gravity in \cite{722}.
Note that the action Eq.(\ref{3.1}) generalizes the Einstein Hilbert action.%
\newline
The choice of an exponential function which couples $R$ with the field $\phi
$, is, in effect, a conformal transformation normalization to conjugate with
standard $f(R)$ gravity. The action Eq.(\ref{3.1}) can be written in the
following form%
\begin{equation}
I=\frac{M_{P}^{2}}{2}\int d^{4}x\sqrt{-g}R\phi (\frac{1}{2}\phi R^{2}+\frac{1%
}{\phi }-R+...)+I_{m},  \label{3.11}
\end{equation}%
The term $R^{2}$ introduces a new spin-zero mode \cite{4} which corresponds
to the scalar field $\phi $ in the bulk. To obtain the kinetic term ($\frac{1%
}{2}\phi R^{2}$), we rewrite the Eq. (\ref{3.11}) by considering the
quantity $R\phi $\ as an angle of transformation. Thus, the lagrangian is
\begin{equation}
\emph{L}(R,\phi )=\frac{M_{P}^{2}}{4}\phi R^{2}-\frac{M_{P}^{2}}{2}(R-\frac{1%
}{\phi }),  \label{ez}
\end{equation}%
such as\ a kinetic energy $T(R,\phi )\equiv \frac{M_{P}^{2}}{4}\phi R^{2}$
is also a non-minimal coupling of the field $\phi $ and $R$. The inverse of
the field $\phi $ is equivalent to a cosmological constant in the context of
($\Lambda CDM)$ model. This allows us to obtain the potential%
\begin{equation}
V(R,\phi )\equiv \frac{M_{P}^{2}}{2}\left( R-\frac{1}{\phi }\right) .
\label{3.3}
\end{equation}%
Using Eq.(\ref{030}), we find
\begin{equation}
\widetilde{V}(R,\phi )=\frac{M_{P}^{2}}{2(1-\phi R)}(R-\frac{R}{1-\phi R}%
)e^{\phi R}.
\end{equation}%
or

\begin{equation}
\widetilde{V}(R,\phi )=\frac{M_{P}^{2}}{2}\mu _{_{\phi R}}(R-\frac{\eta
_{_{\phi R}}}{\phi }),
\end{equation}%
where
\begin{equation}
\mu _{_{\phi R}}=\frac{e^{\phi R}}{1-\phi R}\text{ };\text{ }\eta _{_{\phi
R}}=\frac{\phi R}{1-\phi R},
\end{equation}%
\newline
One can find a similar form of the potential Eq.(\ref{3.3}): $\tilde{V}%
(R,\phi )=e^{\phi R}V(R,\phi )$ with a $e^{\phi R}$ factor in Einstein
frame. This factor represents the rotation transformation from Jordan frame
to Einstein frame, then the term $R\phi $ can be considered as an
transformation angle.\newline

\section{Braneworld $f(R,\protect\phi )$ modified gravity}

In this section, we propose a description of modified gravity in
5-dimensional space-time, such as the bulk is compact. The additional
dimension, which we note $Y,$ is depending on both $R$ and $\phi ,$ which
means that the large extra-dimension changes according to the parameters $R$
and $\phi $. Note that the action Eq.(\ref{3.1}) can describe both the
3-brane and the bulk, only by the above choice of the function of $f(R,\phi
) $. This new technique connecting 4d to 5d by a function $f(R,\phi
)=Re^{-\phi R}$, the transition between 4d and 5d is already covered in \cite%
{SC} will allow us to describe the gravity, simultaneously, on the 3-brane
and on the bulk, by a single term of Lagrangian Eq.(\ref{3.1}). There is
also the standard method of describing the gravity, by summing a term which
describes the gravity in 4d and the other term which describes gravity in 5d
in the Lagrangian. Thus, the action Eq.(\ref{3.1}) can be written in the
following form
\begin{equation}
I=\frac{M_{P}^{2}}{2}\int d^{4}x\sqrt{-g}(R\phi )\frac{1}{\phi }e^{-\phi
R}+I_{m},  \label{3.13}
\end{equation}
The rotation $R\phi $, allows us to go from 3-brane to a five-dimensional
bulk. If this angle is zero, the gravitation will be limited only to
3-brane. In other words, we have only the Einstein's gravity, and the action
(\ref{3.1}) will be the Einstein-Hilbert action. To rid of $R\phi $ in Eq. (%
\ref{3.13}), it is assumed that this term is a result of integration on the
fifth dimension. If we define the parameter $x^{5}=y=LR\phi $ of the fifth
dimension, with $L$ is the radius of the large compact dimension, according
to the model ADD \cite{2},\ we can connect the Planck mass $M_{p}$ in
4-dimensional to Planck mass $M_{5}$ in 5-dimensional, defined by $%
M_{p}^{2}=M_{5}^{3}L.$ Thus the action becomes as
\begin{equation}
I=\frac{M_{5}^{3}}{2}\int_{M_{4}}d^{4}x\sqrt{-g}\frac{1}{\phi }e^{-\phi
R}\int_{0}^{\phi RL}dy+I_{m},  \label{e2}
\end{equation}%
The geometric volume in 3-brane $M_{4}$ is $Vol_{M_{4}}=\int_{M_{4}}d^{4}x%
\sqrt{-g},$ and the standard form volume, in large compact dimension, $Y$ is
$Vol_{Y}=e^{-\phi R}\int_{Y}dy$. We can then define the five-dimensional
bulk as $M_{5}=M_{4}\times Y$, which is defined by metric determinant (a
conformal transformation)
\begin{equation}
g_{(5)}=ge^{-2\phi R},  \label{3r}
\end{equation}%
\newline
In this case, we can write the action (\ref{e2}) in geometrical form by $%
I=Vol_{M_{4}}(\frac{1}{\phi }Vol_{Y})\neq \frac{1}{\phi }%
Vol_{Y}(Vol_{M_{4}}),$ which shows us that the structure of 3-brane depends
exactly on the behavior of the large extra dimensions. This result shows
that our brane is immersed in this large extra dimensions (like in
Randall-Sundrum theory) and not like the very small dimension (like in
Kaluza-Klein theory). We can also say that scalar field $\phi $ lives
outside the brane and the interaction between this field with 3-brane can
generate invisible physical phenomena (DM and DE..). According to Eq. (\ref%
{3r}), we can define%
\begin{equation}
\omega _{\phi }:=-R\phi ,  \label{5}
\end{equation}%
We introduce the FLRW metric $ds^{2}=-dt^{2}+a^{2}(t)d\mathbf{x}^{2}$ of a
flat space, which allows us to write $R=6(\dot{H}+2H^{2})$ to find Eq. (\ref%
{5})\ in the Jordan frame. We then study various Universe eras by
considering scale factor $a\varpropto t^{q},$ where $q$ is the decelaration
parameter. That implies the Hubble parameter is $H=\frac{q}{t}$, so one can
get
\begin{equation}
\phi =\frac{-\omega _{\phi }}{6q(2q-1)}t^{2},  \label{3.22}
\end{equation}%
Since the field $\phi $ depends on the square of time. Then Eq. (\ref{3.22})
led to an accelerating expansion of the Universe or accelerating contraction
of the Universe, according to the parameter $\omega _{\phi }$, then
\begin{equation}
\begin{array}{c}
\omega _{\phi }\prec 0\text{ \ \ }\Longrightarrow \text{ \ }\phi \succ
0\Longrightarrow \text{acceleration,} \\
\omega _{\phi }\succ 0\text{ \ }\Longrightarrow \text{ \ }\phi \prec
0\Longrightarrow \text{contraction.}%
\end{array}
\label{3.23}
\end{equation}%
For a positive value $\left( \omega _{\phi }\succ 0\right) ,$\ the Universe
have a contraction period, which corresponds to a very strong gravitation or
the gravitational effects of dark matter.\newline
The negative value of parameter $(\omega _{\phi }\prec 0)$ implies that the
field $\phi $ exerts a repulsion on the 3-brane, which corresponds exactly
to the accelerating expansion of the Universe. This case represents the dark
energy.\newline
To have a Universe in an accelerating expansion, we must have $\omega _{\phi
}=-R\phi \prec 0$. In the present work, we considere the condition $\omega
_{\phi }=-1.$ Thus the Eq.(\ref{5}) leads to\ $\phi R=1,$ and Eq.(\ref{3.22}%
) is%
\begin{equation}
\phi R=1\Longrightarrow \phi =\frac{t^{2}}{6q(2q-1)},  \label{ZE}
\end{equation}%
We will calculate later all the parameters of the model from this relation.
The field $\phi $ and time are equivalent, and we note that the objective of
Eq (\ref{ZE}) is to show that time play an important role in cosmic history
of the Universe. Recall that, in 4-dimensional $f(R)$ gravity models, it was
shown that the field $\phi =t$ \cite{17}. Here, since we are working on f
(R) gravity in 5-dimensional, we will show that $\phi \sim t^{2}$. Thus, the
scalar field is equivalent to square of time, which is compatible with the
accelerated expansion of the Universe \cite{18}.\newline
Injecting Eq.(\ref{3r}) in Eq.(\ref{e2}), one can find the action decribing
the coupling between the Ricci scalar and the scalar field in 5 dimensions:%
\begin{equation}
I=\frac{M_{5}^{3}}{2}\int_{M_{5}}d^{5}x\sqrt{-g_{(5)}}\frac{1}{\phi }+I_{m}.
\label{3.15}
\end{equation}%
Eq.(\ref{3.15}) shows the existence of a field $\phi $ in 5-dimensional
braneworld. Note that this action depends on the scalar field $\phi ,$ what
is different from the general action of the ref. \cite{7}.

By calculating the variation of action Eq.(\ref{3.15}) we find%
\begin{equation*}
\delta I=\frac{M_{5}^{3}}{2}\int_{M_{5}}d^{5}x\delta (\sqrt{-g_{(5)}})\frac{1%
}{\phi }+\sqrt{-g_{(5)}}\delta (\frac{1}{\phi })+\delta I_{m}
\end{equation*}%
Let us consider that $\delta (\sqrt{-g_{(5)}})=-\frac{1}{2}\sqrt{-g_{(5)}}%
g_{(5)MN}\delta g_{(5)}^{MN}$, which leads to%
\begin{equation*}
\delta I=-\frac{M_{5}^{3}}{2}\int_{M_{5}}d^{5}x\sqrt{-g_{(5)}}\delta
g_{(5)}^{MN}\left( \frac{1}{\phi ^{2}}\frac{\delta \phi }{\delta g_{(5)}^{MN}%
}+\frac{1}{2\phi }g_{(5)MN}\right) +\delta I_{m}
\end{equation*}%
We then obtain the equation of motion of $\phi $%
\begin{equation}
\frac{\delta \phi }{\delta g_{(5)}^{MN}}+\frac{1}{2}g_{(5)MN}\phi =-\frac{1}{%
M_{5}^{3}}\phi ^{2}T_{MN}.  \label{3.17}
\end{equation}%
Eq.(\ref{3.17}) looks like the equation of general relativity, and can
describe well the scalar field $\phi $ in the 5-dimensions. We can simplify
Eq.(\ref{3.17}) by%
\begin{equation}
(g_{(5)}^{MN}\frac{\delta }{\delta g_{(5)}^{MN}}+\frac{5}{2}+\frac{%
g_{(5)}^{MN}T_{MN}}{M_{5}^{3}}\phi )\phi =0,  \label{3.171}
\end{equation}%
$\frac{\delta \phi }{\delta g_{(5)}^{MN}}$ is the scalar field variation on
the manifold $M_{5}$ and $T_{MN}$ is the energy-momentum tensor of matter
given by
\begin{equation}
T_{MN}=\frac{-2}{\sqrt{-g_{(5)}}}\frac{\delta \left( \sqrt{-g_{(5)}}\emph{L}%
_{m}\left[ g^{\mu \nu },\Psi \right] \right) }{\delta g_{(5)}^{MN}},\text{ \
\ \ \ \ \ \ \ \ }\left( M,N=0,1,2,3,5\right)
\end{equation}%
The energy-momentum tensor is defined by the variation of the lagrangian of
matter $\emph{L}_{m}\left[ g^{\mu \nu },\Psi \right] $ in accordance to the
metric $g_{_{(5)MN}}$ on the manifold $M_{5}$(the bulk)$.$ $\Psi $ is the
matter field on 3-brane. \newline
We define a new metric of a flat space, that is written as a direct sum of
Friedmann--Lema\^{\i}tre--Robertson---Walker (FLRW) metric and the fifth
dimension component given by the Eq. (\ref{3r}), with a scale factor $a(t),$
like that
\begin{equation}
ds_{(5)}^{2}=-dt^{2}+a^{2}(t)d\mathbf{x}^{2}+e^{-\phi R}dy^{2}  \label{sss}
\end{equation}%
This metric describes the homogeneous Universe in large structure. We then
multiply $g_{(5)}^{MN}$ to $T_{MN}$ to obtain

\begin{equation}
g_{(5)}^{MN}T_{MN}=g^{\mu \nu }T_{\mu \nu }+g_{(5)}^{55}T_{55};  \label{ttt}
\end{equation}%
Let us consider that $g^{\mu \nu }T_{\mu \nu }=\left( 3\omega -1\right) \rho
_{m}$ \cite{R1}, where $\rho _{m}$ is the energy density of matter and $%
\omega $ is the equation of state of matter.

From Eq.(\ref{sss}), we have $g_{(5)}^{55}=e^{\phi R},$ and one can obtain%
\begin{equation}
g_{(5)}^{MN}T_{MN}=\left( 3\omega -1\right) \rho _{m}+T_{55}e^{\phi R},
\label{3.18}
\end{equation}%
We will consider that the component of the energy-momentum tensor $T_{55}$
describes the dark matter, by the relation
\begin{equation}
T_{55}:=\rho _{DM}  \label{3.19}
\end{equation}%
\qquad From Eq.(\ref{3.171}) and Eq.(\ref{3.18}), one can obtain\newline

\begin{equation}
(g_{(5)}^{MN}\frac{\delta }{\delta g_{(5)}^{MN}}\mathbb{+}\frac{\rho
_{DM}\phi e^{\phi R}}{M_{5}^{3}}+\left( 3\omega -1\right) \frac{\rho
_{m}\phi }{M_{5}^{3}}+\frac{5}{2})\phi =0,  \label{rrr}
\end{equation}%
\newline
To get a equation analogue to the Klein-Gordon equation:\ $\left( \square
+m^{2}\right) \phi =0$, we postulate that:
\begin{equation}
\square :=g_{(5)}^{MN}\frac{\delta }{\delta g_{(5)}^{MN}}+\frac{\rho
_{DM}\phi e^{\phi R}}{M_{5}^{3}},  \label{12}
\end{equation}%
\begin{equation}
m^{2}=\left( 3\omega -1\right) \frac{\rho _{m}\phi }{M_{5}^{3}}+\frac{5}{2}.
\label{13}
\end{equation}

These two relations, Eq.(\ref{12}) and Eq.(\ref{13}), will allow us to
describe the evolution of the density of OM, the density of DM and the
density of DE.

\section{The scalaron mass}

In the previous section, we have introduced a form of correspondence between
the motion equation of the field $\phi $ and the Klein-Gordon equation. This
correspondence will allow us to find the scalar field properties like the
masse $m=m_{\phi }$ of the mass of field $\phi .$ Then, Eq.(\ref{13}) in the
approximation Eq.(\ref{ZE}), can be immediately written%
\begin{equation}
\rho _{m}=M_{5}^{3}\frac{\left( m_{\phi }^{2}-\frac{5}{2}\right) }{t^{2}}%
\frac{6q(2q-1)}{\left( 3\omega -1\right) },  \label{ZE1}
\end{equation}%
\newline
we begin by studing first the Eq.(\ref{13}) in the chronology of the
universe in each era: \newline
\textit{In the radiation-dominated era} $(\omega =\frac{1}{3}):$ we then
find the mass of scalar field as%
\begin{equation}
m_{\phi }=2.22\times 10^{-24}\left[ eV\right] .  \label{masse}
\end{equation}%
This mass satisfies the condition of the current scalaron mass \cite{5}.
Note that the scalar field $\phi $ is a scalaron, we thus calculate the
value of this mass from the radiation-dominated era. If we replace this mass
(Eq.(\ref{masse})) on the density Eq.(\ref{ZE1}), we will have always a
vanishing density matter. To solve this problem, we have used the notion of
chameleonic dark matter \cite{5}, in other words, the mass of the field $%
\phi $ (\ref{masse}) changes over time. If we assume that $\omega $ takes
the following values $\left( -1,0,1/3\right) $, we obtain\newline
\textit{In the matter-dominated era} $(\omega =0),$ one can have
\begin{equation}
\rho _{m}\phi \leq \frac{5M_{5}^{3}}{2},  \label{d1}
\end{equation}%
\newline
\textit{In the dark-energy-dominated era} $\left( \omega =-1\right) $\newline
\begin{equation}
\rho _{m}\phi \leq \frac{5M_{5}^{3}}{8},  \label{d2}
\end{equation}%
\newline
According to\ Eq.(\ref{ZE}), Eq.(\ref{d1}) and Eq.(\ref{d2}), we obtain:%
\newline
\textit{In the radiation-dominated era }$\left( q=\frac{1}{2}\right) $
\begin{equation*}
\rho _{m}=0,
\end{equation*}%
\textit{In the matter-dominated era }$\left( q=\frac{2}{3}\right) $%
\begin{equation}
\rho _{m}\leq \frac{10M_{5}^{3}}{3}\frac{1}{t^{2}},  \label{tt2}
\end{equation}%
This relationship shows us that the maximum value of the energy density of
matter is
\begin{equation}
\rho _{m\left( \max \right) }=\frac{10M_{5}^{3}}{3}\frac{1}{t^{2}}.
\label{TT3}
\end{equation}%
The energy density of matter decreases over the square of time, this is
because of the accelerated expansion of the Universe \cite{18}.\newline

\section{\protect\bigskip A unified model of dark energy and dark matter in
5D}

\subsection{Dark matter density}

In this section, we will study the variations of the dark matter density. To
closely study the dark matter density, we return to Eq.(\ref{12}), and
calculate the density of dark matter with respect to the field $\phi $, we
find%
\begin{equation}
\rho _{DM}=\frac{M_{5}^{3}}{e^{\phi R}\phi ^{2}}\left( \square \phi
-g_{(5)}^{MN}\frac{\delta \phi }{\delta g_{(5)}^{MN}}\right) ,
\end{equation}%
Using the approximation Eq.(\ref{ZE}), we obtain\newline
\begin{equation}
\rho _{DM}=\frac{12q(2q-1)M_{5}^{3}}{et^{2}}\left( -tg_{(5)}^{MN}\frac{%
\delta t}{\delta g^{MN}}+\partial _{\mu }t\partial ^{\mu }t+t\square
t\right) ,  \label{a01}
\end{equation}%
We must first, calculate each term of this density.

From Eq.(\ref{ZE}) and Eq.(\ref{sss}), one can obtain%
\begin{equation}
g_{(5)}^{MN}=\left(
\begin{array}{ccccc}
-1 & 0 & 0 & 0 & 0 \\
0 & a^{-2} & 0 & 0 & 0 \\
0 & 0 & a^{-2} & 0 & 0 \\
0 & 0 & 0 & a^{-2} & 0 \\
0 & 0 & 0 & 0 & e%
\end{array}%
\right) ,  \label{tt0}
\end{equation}%
we have also
\begin{equation}
g_{(5)}^{MN}\frac{\delta }{\delta g_{(5)}^{MN}}=3g^{11}\frac{\delta }{\delta
g^{11}}=\frac{-3}{2}\frac{a\delta }{\delta a},  \label{z5}
\end{equation}%
or, equivalently%
\begin{equation}
g_{(5)}^{MN}\frac{\delta t}{\delta g_{(5)}^{MN}}=\frac{-3a}{2\overset{.}{a}}=%
\frac{-3}{2H}.  \label{p1}
\end{equation}%
On the other hand
\begin{equation}
\partial _{\mu }t\partial ^{\mu }t=-1+\sum_{i=1}^{3}\frac{1}{a^{2}v_{i}^{2}}+%
\frac{e}{v_{y}^{2}},  \label{p2}
\end{equation}%
with $v_{y}$ is the speed of field $\phi $ according to the dimension $Y$ \
and $v_{i\text{ \ }}$are the field speeds according to the dimension of
space. We have also%
\begin{equation}
\square t=0,  \label{p3}
\end{equation}%
From Eq.(\ref{p1}), Eq.(\ref{p2}) and Eq.(\ref{p3}), the expression of the
density of dark matter (\ref{a01}) become
\begin{equation*}
\rho _{DM}=\frac{12q(2q-1)M_{5}^{3}}{et^{2}}\left[ \frac{3t}{2H}%
-1+\sum_{i=1}^{3}\frac{1}{a^{2}v_{i}^{2}}+\frac{e}{v_{y}^{2}}\right] ,
\end{equation*}%
or%
\begin{equation}
\rho _{DM}=\frac{12q(2q-1)M_{5}^{3}}{et^{2}}\left[ \frac{3t}{2H}-1+\frac{1}{%
a^{2}}\frac{1}{v_{x}^{2}}+\frac{1}{a^{2}}\frac{1}{v_{y}^{2}}+\frac{1}{a^{2}}%
\frac{1}{v_{z}^{2}}+e\frac{1}{v_{y}^{2}}\right] ,  \label{z2}
\end{equation}%
From this relation, we can see that the density of dark matter is inversely
proportional to the speed of the Universe's expansion and time, which means,
that the density of dark matter decreases over time. \newline
If we suppose that the Universe is homogeneous and isotropic, then the speed
of cosmic expansion is $v=v_{x}=v_{y}=v_{z}$, in this case we have
\begin{equation}
\rho _{DM}=\frac{12q(2q-1)M_{5}^{3}}{et^{2}}\left[ \frac{3t}{2H}-1+\frac{3}{%
av^{2}}+\frac{e}{v_{y}^{2}}\right] .  \label{dark}
\end{equation}%
Thus, the density of dark matter Eq.(\ref{dark}), depends inversely on the
speed of expansion of the Universe and the time. We observe also that the
density depends inversely on the speed of expansion in the dimension $Y$,
which cannot cancel, in order to have a real density of dark matter. Thus,
it is possible to explain, at the same-time, the cosmic acceleration (Eq.(%
\ref{ZE})) and the choice of the geometry Eq.(\ref{e2}). We can conclude
that, the expansion of the Universe in 4-dimensions is a result of another
expansion of 5-dimensional space-time, and that the large compact dimension $%
Y$ is accelerating expansion.\newline
For $t=0$, we can define the initial speed of the universe's expansion
(3-brane) $v(t=0)=v_{0}$, and the initial speed of the bulk's expansion $%
v_{y}(t=0)=v_{y0}.$ Eq.(\ref{dark}) becomes%
\begin{equation}
\frac{3}{av_{0}^{2}}+\frac{e}{v_{y0}^{2}}=1
\end{equation}%
This equation shows us exactly, the relationship between the initial speed
of the Universe's expansion and the initial speed of the bulk's expansion.
In addition, we cannot determine their values, because this equation is with
three variables ($v_{0},v_{y0},a$). Since the field $\phi $ occupies all the
space in the bulk, we can propose that the kinetic energy of the scalar
field, is responsible for the bulk's expansion, which can be explained the
3-brane's expansion.\newline

\subsection{Matter dominated era\newline
}

In this section, we calculate the scale factor value from the above model.
We consider the matter dominated era $\left( \omega =0,q=\frac{2}{3}\right) $%
. Eq.(\ref{ZE}) leads to
\begin{equation}
\phi =\frac{3t^{2}}{4}\ \newline
,  \label{tt8}
\end{equation}%
Note that, the obtained value of scalar field $\phi $ is equivalent to the
Ricci tensor value \cite{16},\qquad \newline
To study the energy density of the Universe, we use the approximation Eq.(%
\ref{ZE}), in the matter dominated era. Thus%
\begin{equation}
\frac{2\delta t}{\delta g_{(5)}^{MN}}+\frac{g_{(5)MN}}{2}t+\frac{T_{MN}}{%
4M_{5}^{3}}t^{3}=0,
\end{equation}%
We have considered the vacuum without matter $T_{MN}=0$.\newline
Let us consider that $g_{(5)MN}\delta g_{(5)}^{MN}=-$ $g_{(5)}^{MN}\delta
g_{(5)MN},$ to find
\begin{equation}
\frac{\delta \ln t}{\delta x^{\mu }}-\frac{1}{4}g_{(5)}^{MN}\frac{\delta
g_{(5)MN}}{\delta x^{\mu }}=0,
\end{equation}%
The Christoffel symbols is given by: $\Gamma _{\mu N}^{N}=\frac{1}{2}%
g_{(5)}^{MN}\frac{\delta g_{(5)MN}}{\delta x^{\mu }},$ so%
\begin{equation}
\Gamma _{\mu N}^{N}=2\frac{\delta \ln t}{\delta x^{\mu }},  \label{tt}
\end{equation}%
The metric determinat for flat spacetime Eq.(\ref{tt0}) is calculated and we
have%
\begin{equation}
g=-\frac{a^{6}}{e},  \label{i1}
\end{equation}%
Or $\Gamma _{\mu N}^{N}=\frac{1}{2}\frac{\delta \ln g}{\delta x^{\mu }}\ $%
and \ according to Eq.(\ref{tt}), one can find%
\begin{equation}
g=g_{0}t^{4},  \label{i2}
\end{equation}%
We compare the two equations Eq.(\ref{i1}) and Eq.(\ref{i2}), and they
correspond to
\begin{equation}
a^{6}=-eg_{0}t^{4},
\end{equation}%
we obtaine then the scale factor%
\begin{equation}
a(t)=\left( -eg_{0}\right) ^{\frac{1}{6}}t^{\frac{2}{3}},  \label{tt9}
\end{equation}%
This relationship is consistent with the results of $a\propto t^{\frac{2}{3}%
} $ \cite{16}, and this model can have cosmological interpretations. We
solve now the Eq.(\ref{3.17}) in an Universe without matter and without dark
matter $\left( T_{MN}=0\right) ,$ according to the metric Eq.(\ref{tt0}),
and we find%
\begin{equation}
\frac{\delta \phi }{\phi }=\frac{3\delta a}{a},
\end{equation}%
We then immediately obtain the following relations%
\begin{equation}
a(t)=\frac{\chi }{\left( 6q\left( 2q-1\right) \right) ^{\frac{1}{3}}}t^{%
\frac{2}{3}}.  \label{AZ}
\end{equation}%
with $\chi $ is the integration constant.\newline
We observe that, even if we chose the condition of a Universe without matter
$\left( T_{MN}=0\right) $ and by the general equation Eq.(\ref{3.17}), we
find also the result $a\propto t^{\frac{2}{3}}$. This comes from the choice
of metric; Eq.(\ref{tt0}) and the condition; Eq.(\ref{ZE}). These two
conditions describes well the matter-dominated era. \newline

\subsection{Dark energy dominated era}

We study now the field properties in the dark energy dominated era. Since
this field represents the accelerating expansion of the Universe, we propose
that the values $\phi \longrightarrow \infty $ are equivalent to the dark
energy dominated era.\newline
We now propose to solve Eq.(\ref{3.17}) without the condition Eq.(\ref{ZE})
\begin{equation}
\frac{g_{(5)}^{MN}\delta }{\delta g_{(5)}^{MN}}\frac{1}{\phi }-\frac{5}{%
2\phi }=\frac{1}{M_{5}^{3}}g_{(5)}^{MN}T_{MN}
\end{equation}%
Using Eq.(\ref{z5}) and Eq.(\ref{3.18}), we obtain%
\begin{equation}
\frac{\left( 3\omega -1\right) }{M_{5}^{3}}\rho _{m}+\frac{1}{M_{5}^{3}}%
e^{\phi R}\rho _{DM}+\frac{3}{2\phi ^{2}}\frac{a\delta \phi }{\delta a}=%
\frac{5}{2\phi }  \label{ss1}
\end{equation}%
Consider now the dark-energy-dominated era $\left( \omega =-1\right) $, by
assuming an acceleration which tends towards infinity, in other words $\phi
\longrightarrow \infty $. In this case, Eq.(\ref{ss1}) reduces to%
\begin{equation}
\phi =\frac{1}{R}\ln (\frac{4\rho _{m}}{\rho _{DM}})  \label{ss2}
\end{equation}%
\newline
Note that, one can obtain the equation (\ref{ZE}) from Eq(\ref{ss2}) if $%
\rho _{DM}=\frac{4}{e}\rho _{m}.$ \newline
In 3-brane model, we propose that the scalar field is zero, this indicate
that $\rho _{DM}=4\rho _{m}$, and this result is almost equivalent to the
percentages of the density of dark matter (DM) and the density of ordinary
matter (OM) in the Universe \cite{6}: $\rho _{DM}=80\%$ \ ; $\rho _{m}=20\%.$
On the other hand, since $\phi \longrightarrow \infty ,$ this is equivalent
to $\rho _{DM}\prec \prec 4\rho _{m}$. This result indicate that, when the
acceleration increases, the density of dark matter decreases, which means
that, the mass of the field $\phi $ transforms to kinetic energy, and we can
suggest then that dark matter can be transformed to dark energy.

\section{Conclusion}

In this paper, we have studied the evolution of the scalar field $\phi $ on
the 3-brane and in the bulk, through a passage that transforms all
parameters of 4-dimensional model to 5-dimensional one. The 5-dimensional
theory depends on the field $\phi $. We have shown that the description of
dark energy, dark matter and ordinary matter is obtained by the evolution
equation of the scalar field $\phi $. We found a relation between the
density of DM and the speed of the universe's expansion. The expansion speed
on the 3-brane depends on the expansion speed of bulk. Note that the mass of
field $\phi $ creates dark matter, which changes over time. This is called
chameleonic dark matter. On the other hand, we have shown that, the kinetic
energy of scalar field, is responsible for the bulk's expansion, which means
that, the bulk generates an the accelerating expansion of the Universe. The
decrease of the scalar field mass with the acceleration of Universe
expansion, shows that the mass of the field $\phi $ is transformed to an
energy (dark energy). \newline


\begin{thebibliography}{99}
\bibitem{R1} Amendola, L., Polarski, D., \& Tsujikawa, S. (2007). \textrm{%
Are f (R) dark energy models cosmologically viable?.} \emph{Physical review
letters, 98(13), 131302.}

\bibitem{R2} Sotiriou, T. P., \& Faraoni, V. (2010). \textrm{f (R) theories
of gravity. }\emph{Reviews of Modern Physics, 82(1), 451.}

\bibitem{R3} Odintsov, S. D., G\'{o}mez, D. S. C., \& Sharov, G. S. (2019).
\textrm{Testing logarithmic corrections to R 2-exponential gravity by
observational data.} \emph{Physical Review D, 99(2), 024003.}

\bibitem{R4} Avilez, A., \& Skordis, C. (2014). \textrm{Cosmological
constraints on Brans-Dicke theory. }\emph{Physical review letters, 113(1),
011101.}

\bibitem{R5} Flanagan, E. E. (2004). \textrm{Palatini form of 1/R gravity. }%
\emph{Physical review letters, 92(7), 071101.}

\bibitem{R6} Amendola, L., Gannouji, R., Polarski, D., \& Tsujikawa, S.
(2007). \textrm{Conditions for the cosmological viability of f (R) dark
energy models.} \emph{Physical Review D, 75(8), 083504.}

\bibitem{4} Cembranos, J. A. (2009). \textrm{Dark matter from R 2 gravity.}
\emph{Physical review letters, 102(14), 141301.}

\bibitem{71} Farajollahi, H., Setare, M., Milani, F., \& Tayebi, F. (2011).
\textrm{Cosmic dynamics in }$f(R,\phi )$\textrm{\ gravity.} \emph{General
Relativity and Gravitation, 43(6), 1657-1669.}

\bibitem{711} Capozziello, Salvatore, et al. \textrm{Constraining extended
gravity models by S2 star orbits around the Galactic Centre.} \emph{Physical
Review D 90.4 (2014): 044052.}

\bibitem{722} Stabile, A., and S. Capozziello. \textrm{Galaxy rotation
curves in}\ $f(R,\phi )$ \textrm{gravity}. \emph{Physical Review D 87.6
(2013): 064002.}

\bibitem{73} Zhong, Y., Liu, Y. X., \& Yang, K. (2011). \textrm{Tensor
perturbations of f (R)-branes.} \emph{Physics Letters B, 699(5), 398-402.}

\bibitem{7} Cui, Z. Q., Liu, Y. X., Gu, B. M., \& Zhao, L. (2018). \textrm{%
Linear stability of }$\mathrm{f(R,}\phi \mathrm{,X)}$\textrm{\ thick branes:
tensor perturbations.} \emph{Journal of High Energy Physics, 2018(11), 83.}

\bibitem{10} Randall, L., \& Sundrum, R. (1999). \textrm{Large mass
hierarchy from a small extra dimension.} \emph{Physical review letters,
83(17), 3370}.

\bibitem{9} Gu, B. M., Liu, Y. X., \& Zhong, Y. (2018). \textrm{Stable
Palatini f (R) braneworld.} \emph{Physical Review D, 98(2), 024027.}

\bibitem{13} Zarrouki, R., \& Bennai, M. (2010). \textrm{Chaplygin gas
braneworld inflation according to WMAP7 data.} \emph{Physical Review D,
82(12), 123506.}

\bibitem{SCA} Gannouji, R., Sami, M., \& Thongkool, I. (2012). \textrm{%
Generic f (R) theories and classicality of their scalarons. }\emph{Physics
Letters B, 716(2), 255-259.}

\bibitem{SC} Capozziello, Salvatore, Giuseppe Basini, and Mariafelicia De
Laurentis. \textrm{Deriving the mass of particles from Extended Theories of
Gravity in LHC era.} \emph{The European Physical Journal C 71.6 (2011):
1679. }

\bibitem{2} Antoniadis, I., Arkani-Hamed, N., Dimopoulos, S., \& Dvali, G.
(1998). \textrm{New dimensions at a millimeter to a Fermi and superstrings
at a TeV.} \emph{Physics Letters B, 436(3-4), 257-263.}

\bibitem{MG} Nojiri, S. I., \& Odintsov, S. D. (2007). \textrm{Introduction
to modified gravity and gravitational alternative for dark energy.} \emph{%
International Journal of Geometric Methods in Modern Physics, 4(01), 115-145.%
}

\bibitem{6} P.Collaboration, P. A. R.Ade, , N.Aghanim, , \& C.
Armitage-Caplan, (2014). \emph{Planck 2013 results. XVI. Cosmological
parameters.} \textrm{Astron. Astrophys, 571, A16.}

\bibitem{16} AA. Starobinsky, (2007). Disappearing cosmological constant in
f(R) gravity. Springer, JETP letters.

\bibitem{17} Setare, M. R. \emph{"Holographic modified gravity."} \textrm{%
International Journal of Modern Physics D 17.12 (2008): 2219-2228.}

\bibitem{18} Riess, Adam G., et al. \emph{"Observational evidence from
supernovae for an accelerating universe and a cosmological constant."}
\textrm{The Astronomical Journal 116.3 (1998): 1009.}

\bibitem{5} TKatsuragawa, T., \& Matsuzaki, S. (2018). \textrm{Cosmic
history of chameleonic dark matter in F (R) gravity.} \emph{Physical Review
D, 97(6), 064037.}
\end{thebibliography}
\end{document}